\documentclass{JINST}
\let\ifpdf\relax

\usepackage{lineno}
\usepackage{subfigure}
\usepackage{comment}

\usepackage{siunitx}

\usepackage{placeins}

\graphicspath{{./images/}}

\title{Characterisation of novel prototypes of monolithic HV-CMOS pixel detectors for high energy physics experiments}

\author{S.~Terzo$^a$\thanks{Corresponding author.}~, E.~Cavallaro$^a$, R.~Casanova$^a$,  F.~Di~Bello$^b$, F.~F\"orster$^a$, S.~Grinstein$^{a,c}$, I.~Per\'ic$^d$, C.~Puigdengoles$^a$, B.~Ristic$^{b,e}$, M.~Vicente~Barrero~Pinto$^b$ and E.~Vilella$^f$\\
\llap{$^a$}Institut de F\'isica d'Altes Energies (IFAE), The Barcelona Institute of Science and Technology,
 Edifici CN, UAB campus, 08193 Bellaterra (Barcelona), Spain\\
  \llap{$^b$} D\'epartement de Physique Nucl\'eaire et Corpusculaire (DPNC), University of Geneva,
24 quai Ernest Ansermet, 1211 Genève 4, Switzerland\\
 \llap{$^c$}Instituci\'o Catalana de Recerca i Estudis Avan\c{c}ats (ICREA),
Pg. Llu\'is Companys 23, 08010 Barcelona, Spain\\
\llap{$^d$}Karlsruher Institut f\"ur Technologie (KIT),
 Keiserstra{\ss}e 12, 76131 Karlsruhe, Germany\\
\llap{$^e$} Conseil Europ\'een pour la Recherche Nucle\'aire (CERN),
385 route de Meyrin, 1217 Meyrin, Switzerland\\
 \llap{$^f$}University of Liverpool,
Oliver Lodge, Oxford Street, Liverpool L69 7ZE, United Kingdom\\
E-mail: \email{Stefano.Terzo@cern.ch}}

\abstract{An upgrade of the ATLAS experiment for the High Luminosity phase of LHC is planned for 2024 and foresees the replacement of the present Inner Detector (ID) with a new Inner Tracker (ITk) completely made of silicon devices. Depleted active pixel sensors built with the High Voltage CMOS (HV-CMOS) technology are investigated as an option to cover large areas in the outermost layers of the pixel detector and are especially interesting for the development of monolithic devices which will reduce the production costs and the material budget with respect to the present hybrid assemblies. For this purpose the H35DEMO, a large area HV-CMOS demonstrator chip, was designed by KIT, IFAE and University of Liverpool, and produced in AMS 350 nm CMOS technology. It consists of four pixel matrices and additional test structures. Two of the matrices include amplifiers and discriminator stages and are thus designed to be operated as monolithic detectors. In these devices the signal is mainly produced by charge drift in a small depleted volume obtained by applying a bias voltage of the order of \SI{100}{V}. Moreover, to enhance the radiation hardness of the chip, this technology allows to enclose the electronics in the same deep N-WELLs which are also used as collecting electrodes.
In this contribution the characterisation of H35DEMO chips and results of the very first beam test measurements of the monolithic CMOS matrices with high energetic pions at CERN SPS will be presented.}

\begin{document}

\section{Introduction}\label{sec:intro}
The HV-CMOS technology is being investigated for the Inner Tracker (ITk) of the ATLAS detector upgrade at the High Luminosity Large Hadron Collider (HL-LHC) as an option for the fifth pixel barrel layer~\cite{sterzo}. A design with a large fill factor consisting of transistors embedded in the same deep N-WELL used as collecting electrode, first proposed by I.~Per\'ic in Ref.~\cite{peric1}, is being investigated as potentially radiation hard. This sensor technology has already shown good results in terms of radiation hardness in Capacitive Coupled Pixel Detectors (CCPDs)~\cite{h18irr}. To continue this line of research, a large scale demonstrator which also includes standalone monolithic structures has been developed. indeed, Depleted Monolithic Active Pixel Sensors (DMAPS) would be an attractive cost-effective alternative to the present hybrid planar pixel module technology for the outer pixel layers of the ITk. In the following, characterisation and first beam test measurements on the monolithic part of a novel HV-CMOS chip developed for ITk studies are presented.

\section{The H35 demonstrator}
The H35DEMO~\cite{h35demo} is a large area HV-CMOS demonstrator chip produced in \SI{350}{nm} AMS\footnote{Austria Mikro Systeme, \href{http://ams.com}http://ams.com} technology on different wafer resistivities: \SIlist{20;80;200;1000}{\ohm cm}. It was designed by a collaboration of KIT, IFAE and University of Liverpool. The chip is composed of four matrices: two analog matrices of \SI{300x32}{} pixels with in-pixel amplification meant to be capacitive coupled to ATLAS FE-I4 readout chips, and two monolithic matrices of \SI{300x16}{} pixels provided with a digital periphery part for standalone operations. For all matrices the pixel size is \SI{50x250}{\mu m}. The first monolithic matrix, referred to as nMOS matrix, contains digital pixels with in-pixel Charge Sensitive Amplifiers (CSAs), shapers and nMOS discriminators embedded in the same deep N-WELL acting also as collecting electrode. The output of the discriminator of each pixel is then connected to an additional CMOS discriminator inside the Readout Cells (ROCs) which are placed in the chip periphery. In the second monolithic matrix, referred to as CMOS matrix, the pixel structure is the same, but does not include any in-pixel discriminator. The analog signals from the shapers are thus directly transmitted to the CMOS discriminators in the periphery. Each monolithic matrix is divided into two sub-matrices. The right nMOS sub-matrix contains an additional in-pixel discriminator with time-walk compensation, while in the right CMOS sub-matrix a second discriminator is present in the digital periphery which generates an additional time stamp. The ROCs are identical in the two monolithic matrices and are arranged in 120 columns (60 for each sub-matrix) and 40 rows read out with a column drain architecture. 
Each ROC column is terminated with an End Of Column (EOC) logic which registers the time stamp and the address of the pixels. The content of each EOC is processed sequentially by a control unit. The readout of the matrix is asynchronous with no trigger. A serialiser for each sub-matrix transmits the data off-chip with a frequency as high as \SI{320}{MHz}. Additional test structures without electronics are included in the chip for capacitance and Transient Current Technique (TCT) measurements. A full TCT characterisation of the latter test structures on different resistivity substrates before and after irradiation can be found in Ref.~\cite{ecavalla}.

\section{The IFAE readout system}\label{sec:readout}
\begin{figure*}[tbph!] 
\centering
	\includegraphics[width=.54\textwidth]{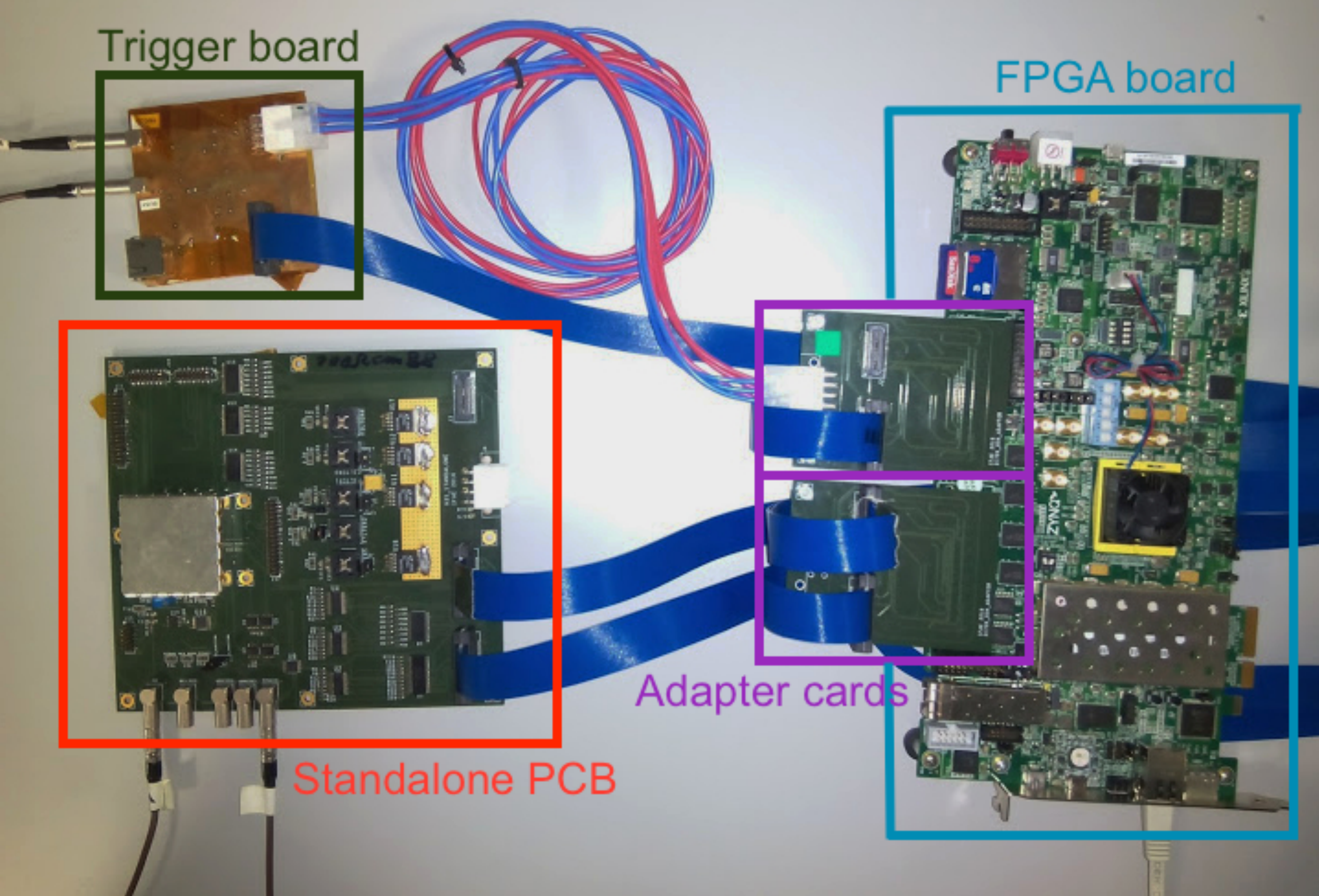}
\caption{The IFAE readout system for the monolithic matrices of the H35DEMO chip. The setup shown in the picture includes the FPGA board with two adapter cards connected, the trigger board and the first version of the standalone PCB.}
\label{fig:readout_system}
\end{figure*}

A readout system to program and operate the two monolithic matrices of the H35DEMO has been developed at IFAE (Figure~\ref{fig:readout_system}). It is based on a Xilinx ZC706 FPGA kit\footnote{Xilinx, Inc.: \href{https://www.xilinx.com}https://www.xilinx.com} connected to a standalone Printed Circuit Board (PCB) on which the H35DEMO chip is glued and both the nMOS and the CMOS monolithic matrices are wire-bonded. The FPGA board is connected via ethernet to a laptop running a steering software written in C++. 
Different adapter cards have been designed for connection with flexible cables on the ZC706 FPGA board: one enables connections with the standalone PCB for communications with the chip and another gives access to the chip test signals. The letter can be substituted by a third adapter card connected to a trigger board which provides input and output TTL signals to be used in beam test setups.
The standalone PCB includes a DAC7568 to regulate the voltages for the chip operations. The bias voltage for the sensor is instead delivered by an independent external power source. An external injection pulse can be also provided with a commercial waveform generator and can be directed to the nMOS or to the CMOS matrix. Connections to monitor the output signals of the CSAs of both matrices are also present on this board.
A first version of the standalone PCB was produced which implements all the connections for powering, operating and testing all the functionalities of the two monolithic matrices of the H35DEMO chip. To correctly operate the chip both monolithic matrices need to be power and configured. Nevertheless, in this configuration a high leakage current, of the order of milliamps, is observed when applying the sensor bias (which is common to the whole chip). A second version of the PCB has been recently design to power and configure also the two analog matrices of the demonstrator. This new PCB reduced the leakage current of the sensor to the order of micro amperes. For the measurements presented in the following the first version of the PCB was used, since the revised version was not yet developed. All the results shown in this contribution have been obtained with the CMOS matrix.

\begin{figure*}[tbph!] 
\centering
\subfigure[Test injection scan]{
	\includegraphics[width=.47\textwidth]{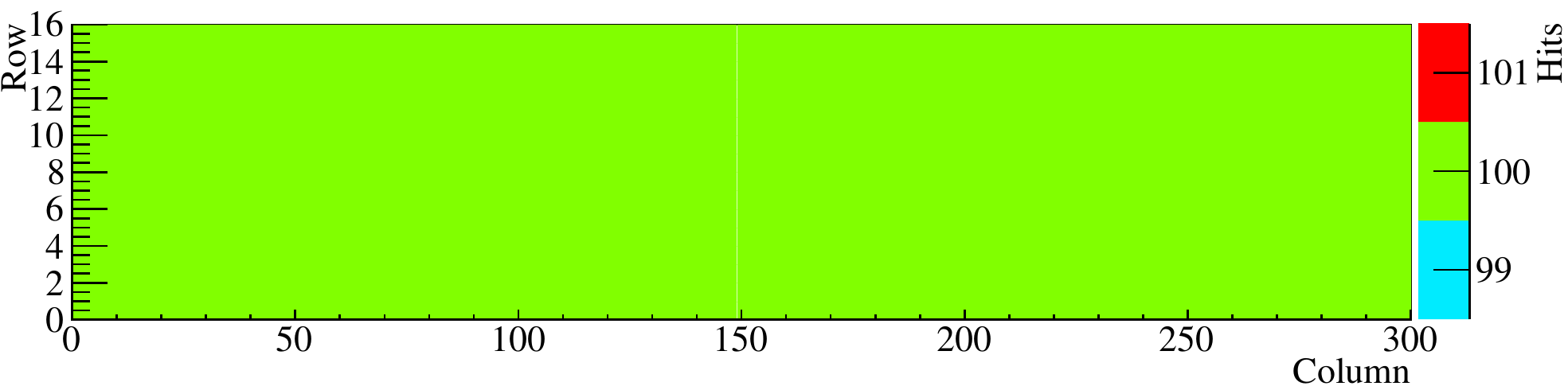}
	\label{fig:analogscan}
}
\subfigure[Source scan]{
	\includegraphics[width=.47\textwidth]{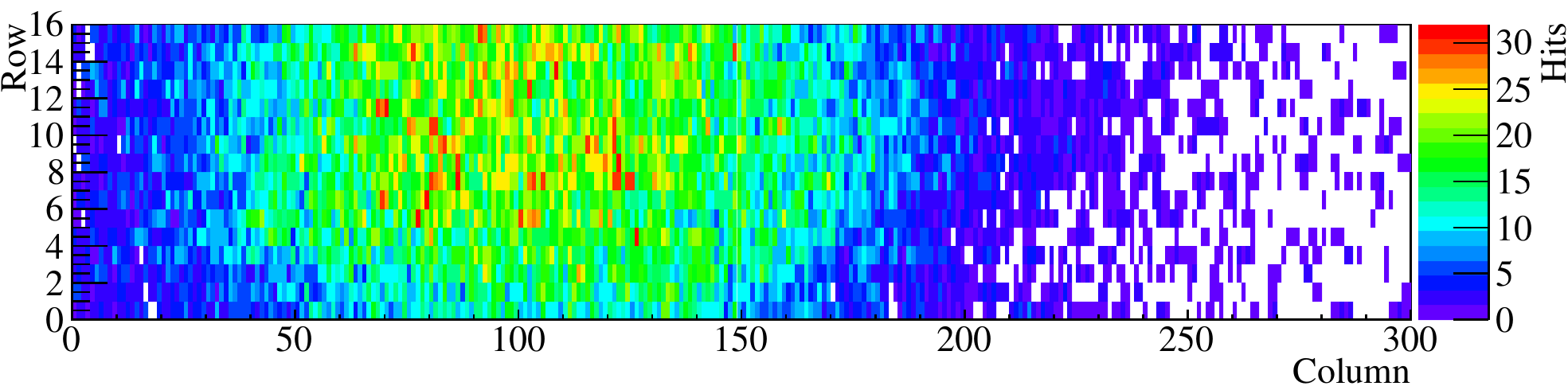}
	\label{fig:sourcescan}
}
\caption{Test of the functionalities of the CMOS monolithic matrix. In \protect\subref{fig:analogscan} and  \protect\subref{fig:sourcescan} the response over the matrix surface to a test injection pulse and $^{90}$Sr beta electrons are shown, respectively.}
\label{fig:scans}
\end{figure*}

\paragraph{Functionality tests}\label{sec:test}
The FPGA and the chip are both operated at \SI{320}{MHz} and each EOC containing pixel addresses and time stamps is read out from the chip by the FPGA at \SI{40}{MHz}. An injection scan is performed to test the digital response of the matrix. Each pixel is injected with 100 square pulses of \SI{100}{ns} width and \SI{1.5}{V} amplitude at \SI{2}{kHz} and the number of detected hits is checked. The result is a uniform map over the chip surface as shown in Figure~\ref{fig:analogscan}. Response to charged particles is also measured using a $^{90}$Sr beta source. In the result shown in Figure~\ref{fig:sourcescan} the spot of the source due to the collimator can be clearly seen.

\begin{figure*}[tbph!] 
\centering
\subfigure[\SI{0.87}{V}]{
	\includegraphics[width=.31\textwidth]{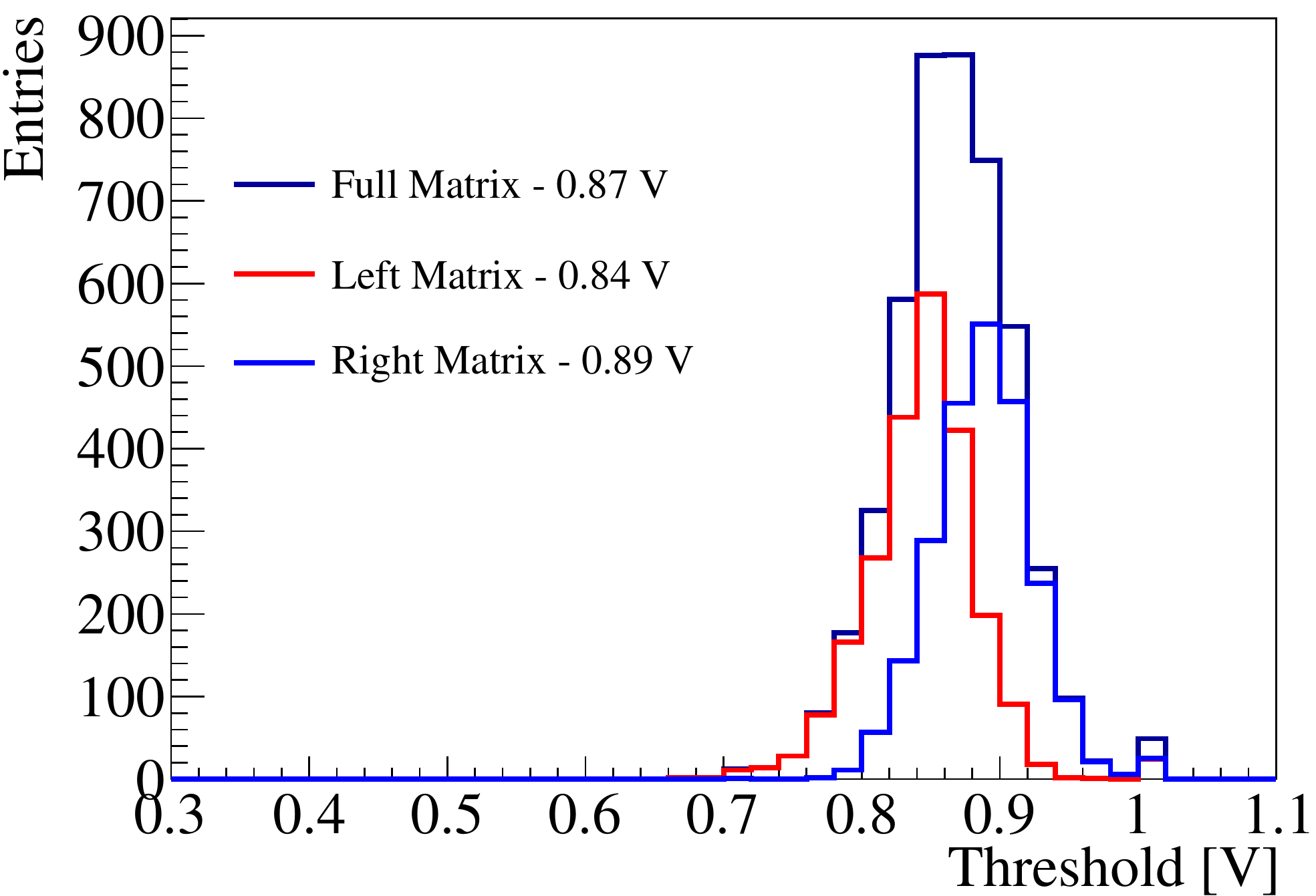}
	\label{fig:thrhigh}
}
\subfigure[\SI{0.71}{V}]{
	\includegraphics[width=.31\textwidth]{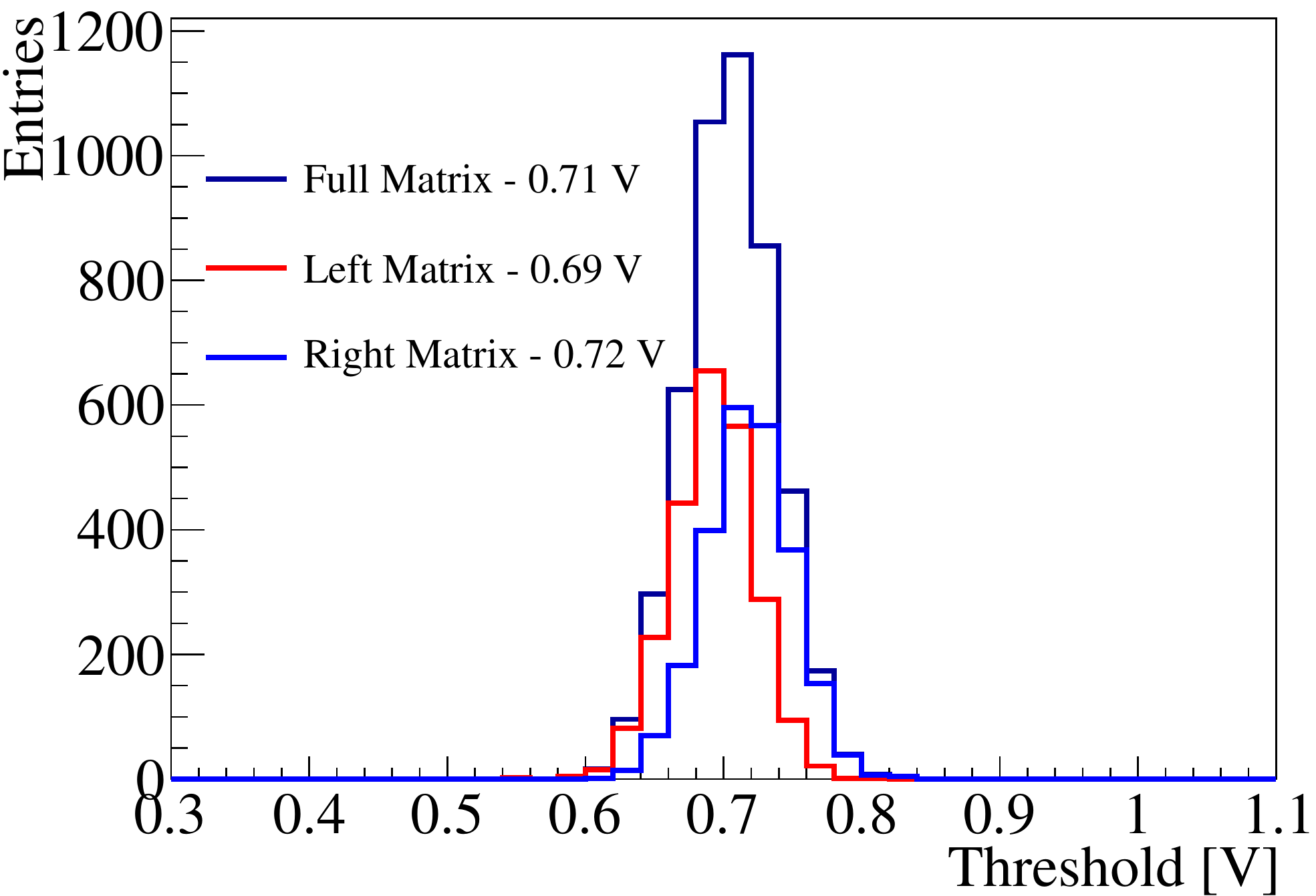}
	\label{fig:thrmed}
}
\subfigure[\SI{0.54}{V}]{
	\includegraphics[width=.31\textwidth]{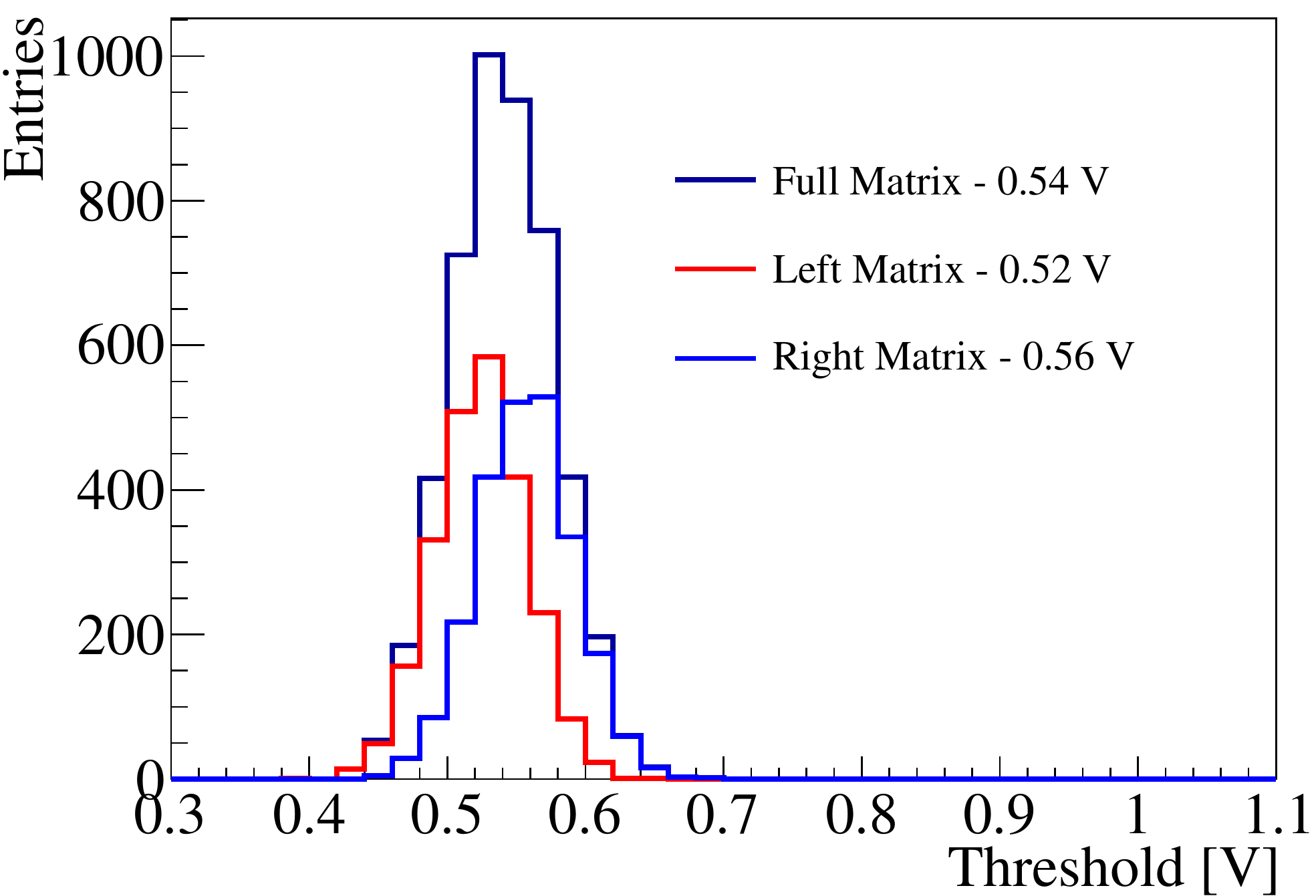}
	\label{fig:thrlow}
}
\subfigure[Threshold map]{
	\includegraphics[width=.56\textwidth]{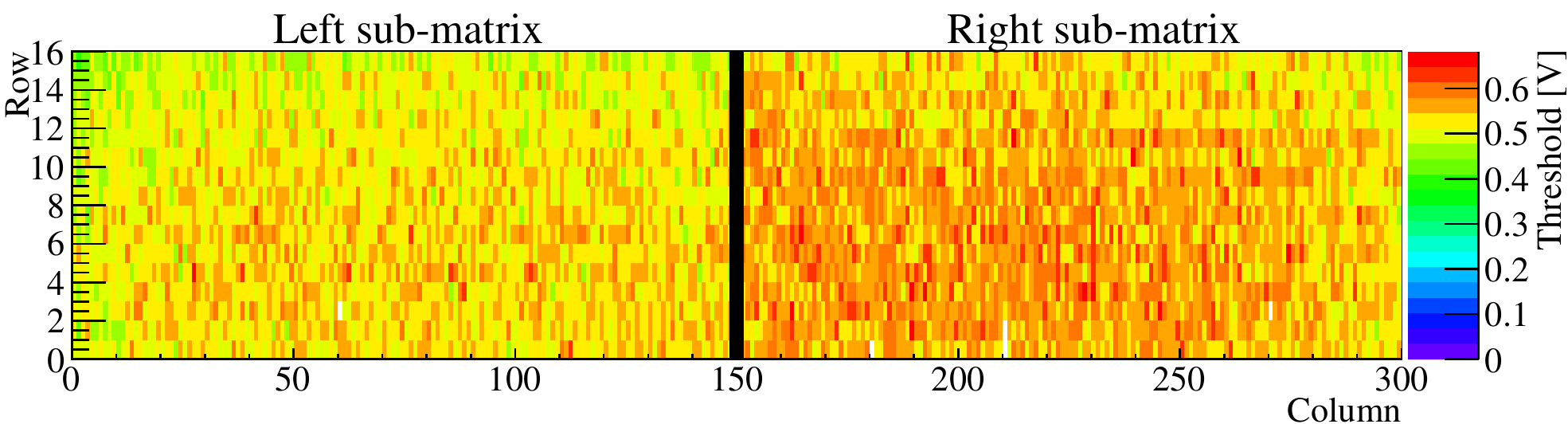}
	\label{fig:thrlowmap}
}
\caption{Threshold tuning of the CMOS matrix of a \SI{200}{\ohm cm} H35DEMO chip. The threshold distributions for three different settings of the global discriminator are shown in \protect\subref{fig:thrhigh}, \protect\subref{fig:thrmed} and \protect\subref{fig:thrlow} resulting in a mean value of \SIlist{0.87;0.71;0.54}{V}, respectively. The distribution of the entire matrix and the corresponding mean value are shown in black, while distributions and mean values for the left and the right sub-matrices are shown in red and blue, respectively. In \protect\subref{fig:thrlowmap} a map representation of the threshold over the matrix surface corresponding to \protect\subref{fig:thrlow} is shown. The units of the threshold are relative to the amplitude of the injected pulse.}
\label{fig:thr}
\end{figure*}

\paragraph{Module tuning}\label{sec:tuning}
The threshold in the CMOS matrix is modified for all pixels by changing the external voltages provided globally to the discriminators. The threshold of each pixel can be additionally adjusted though dedicated on-chip DAC registers of 4-bits, but for the measurements here presented this feature was not exploited.
The resulting threshold is measured for each pixel injecting pulses with different amplitudes. In each pixel 100 injections per amplitude are performed and the threshold point corresponding to \SI{50}{\%} occupancy 
is determined by interpolation with a Gauss error function. For the beam test measurements described in the following section, the chips are tuned to three different thresholds, the corresponding distributions are shown in Figure~\ref{fig:thr}. For all tunings the threshold spread over the pixel surface was about \SI{40}{mV} with a noise of about \SI{30\pm3}{mV}. A systematically higher average threshold was observed in the right sub-matrix probably because of the different discriminator structure which could lead to different threshold for the same parameter settings without a fine tune using the 4-bit DACs.

\section{Beam test measurements}\label{sec:bt}
\begin{figure*}[tbph!] 
\centering
\subfigure[Average trigger time stamp]{
	\includegraphics[width=.51\textwidth]{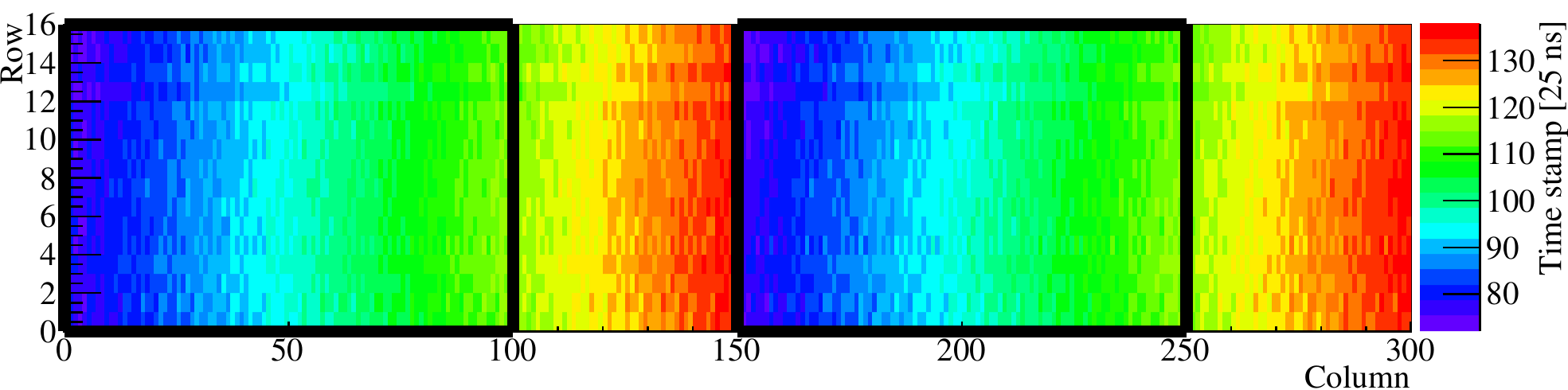}
	\label{fig:tsmap}
}
\subfigure[Hit map]{
	\includegraphics[width=.51\textwidth]{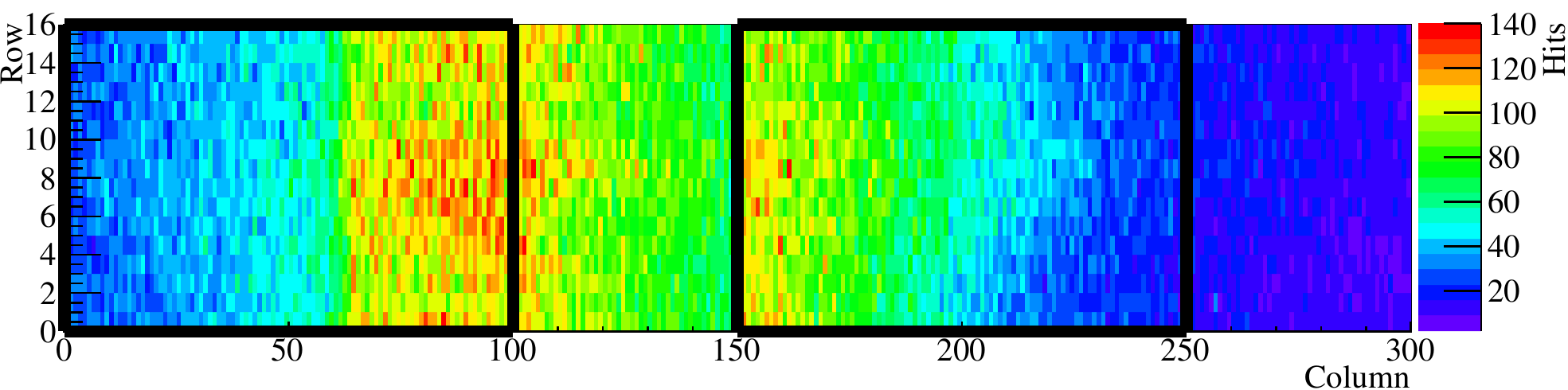}
	\label{fig:hitmap}
}
\subfigure[Hit efficiency]{
	\includegraphics[width=.51\textwidth]{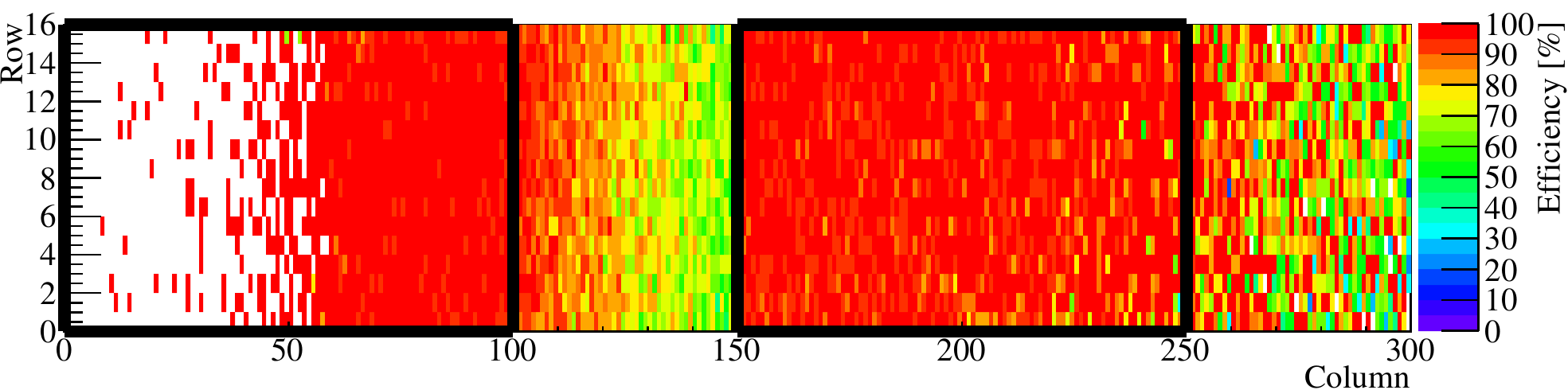}
	\label{fig:effmap}
}
\caption{\protect\subref{fig:tsmap} shows the average time in units of LHC bunch crossings (\SI{1}{bx} = \SI{25}{ns}) that the hit information of the different pixels needs to reach the FPGA after a trigger is received. The FWHM of the flat distribution is constant for all pixels and is \SI{60}{bx}. The black square indicates the Regions Of Interest (ROIs) selected for the analysis. The same ROIs are marked on the hit map in \protect\subref{fig:hitmap} and on the corresponding efficiency map in \protect\subref{fig:effmap} which were taken at the beam test with a limited trigger window. 
The sensor was operated at \SI{120}{V} with a mean threshold of \SI{0.54}{V}.}
\label{fig:maps}
\end{figure*}

\begin{figure*}[tbph!] 
\centering
	\includegraphics[width=.49\textwidth]{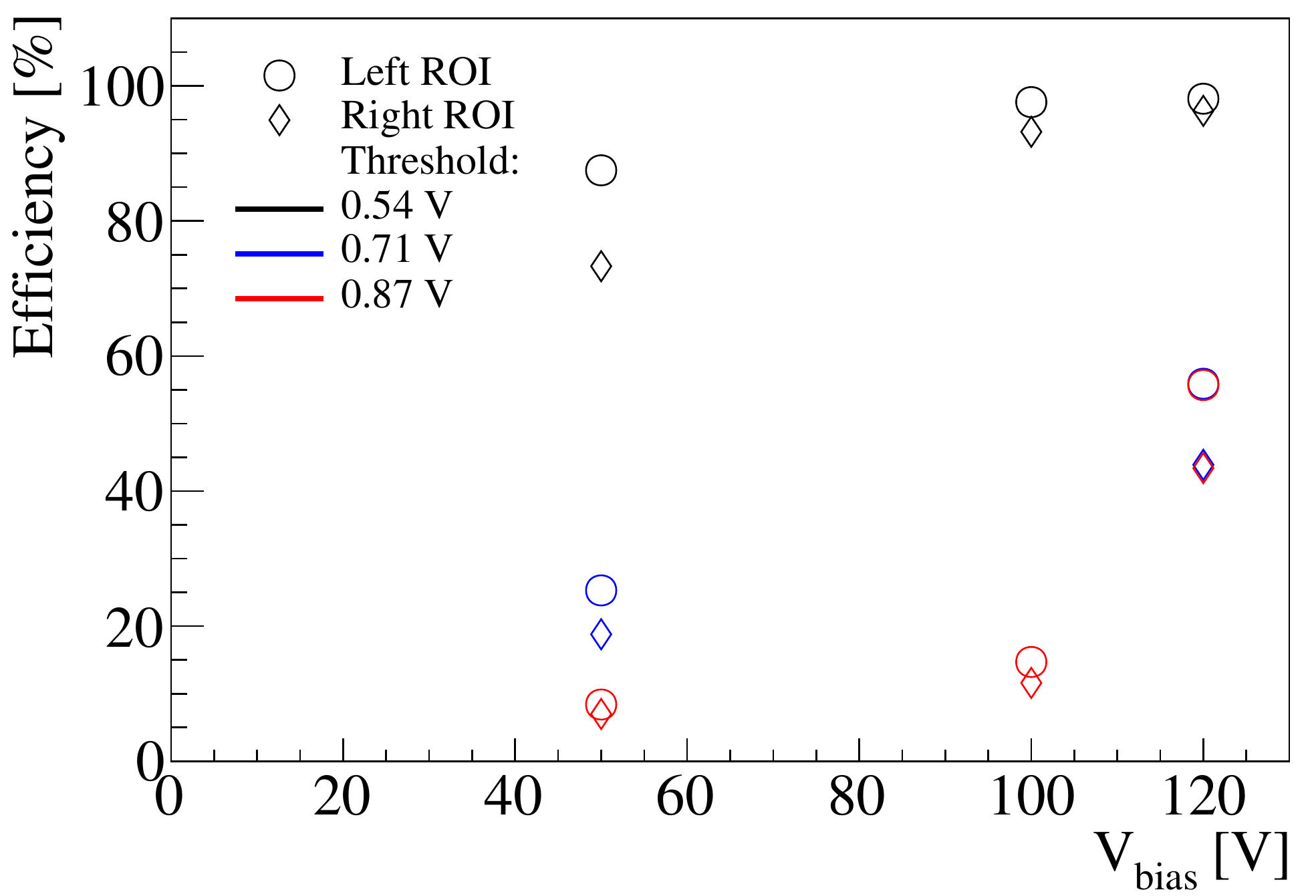}
\caption{Hit efficiency as function of the bias voltage of the monolithic CMOS matrix of a \SI{200}{\ohm cm} H35DEMO chip. The chip was tuned to three different thresholds as shown in Figure~\ref{fig:thr}. The efficiencies are calculated separately for the left and right sub-matrices inside the ROI defined in Figure~\ref{fig:maps}.}
\label{fig:eff}
\end{figure*}

The CMOS matrix of a \SI{200}{\ohm cm} H35DEMO chip was characterised using \SI{180}{GeV} pions in the H8 beam line at CERN SPS.
For particle tracking the UniGe telescope~\cite{fei4telescope} composed of six ATLAS FE-I4 planar modules and based on the RCE readout system~\cite{rce} was used. The IFAE readout system is synchronised with the telescope through a busy-signal scheme: the trigger signal, generated by the coincidence of the first and last planes of the telescope, is transmitted to the IFAE readout system which replies with a busy signal inhibiting the triggering of the telescope until the full system is again ready to receive and process the next trigger event. The data is written sequentially by the separate readout systems and includes a time stamp associated to each event (i.e. trigger and corresponding hit information). This allowed for re-synchronisation in case trigger signal losses occurred. 
For the trigger association to the particle and to reduce the dead time, a limited time window is defined in which the data received from the H35DEMO chip is recorded after a trigger signal is received. To associate the trigger to the particle hit in the H35DEMO chip, and to reduce the dead time, a limited time window is defined in which the data transmitted from the chip is recorded after the trigger signal is received. Each EOC in the digital periphery is read out in a clock cycle of \SI{25}{ns} and the shift registers of the two sub-matrices, consisting of 60 EOCs each, are processed in parallel. Thus the time needed to read out the full ROCs is of about \SI{1.5}{\mu s}. As shown in Figure~\ref{fig:tsmap} this also means that the information from the right columns of each sub-matrix arrives on average later to the FPGA than the one of the left columns.
Due to the continuous readout architecture an additional delay of up to a maximum of \SI{1.5}{\mu s} is expected to read the previous shift register to the one containing the information relative to the received trigger. Since the trigger window was limited during the measurements, a hit loss affecting the information of the rightmost columns of each of the two sub-matrices was experienced. The effect is evident in the central part of the hit map shown in Figure~\ref{fig:hitmap} where the shape of the beam spot is altered by a hit deficit. In the same map the sharp cut of the trigger planes and the periphery of the beam spot can be seen on the leftmost and on the rightmost parts, respectively. The result is a lower efficiency localised in rightmost parts of each sub-matrix as shown in Figure~\ref{fig:effmap}. The analysis was thus restricted to the Region Of Interest (ROI) marked by the black rectangles in Figure~\ref{fig:maps}. These are chosen to exclude the pixels affected by the trigger window limitations of the readout system. 
The summary of the hit efficiencies measured as a function of the applied bias voltage and for different threshold settings is shown in Figure~\ref{fig:eff}. The two higher thresholds chosen resulted to be both too high to operate the chip with a hit efficiency higher than \SI{60}{\%} even with a bias voltage of \SI{120}{V}. While with the lower threshold settings  the hit efficiency reach its maximum of about \SIrange{97}{98}{\%} at \SI{120}{V}. A difference is also observed in the performance of the right and the left sub-matrices with the former one showing consistently higher efficiency for all bias voltages. This can be addressed to the difference in the average threshold between the two sub-matrices shown in Figure~\ref{fig:thr}. The effect is indeed larger as the bias voltage decreases and lower signals, due to the smaller depleted thickness, have more probability to fall below the threshold. In the left sub-matrix, i.e. the one with the lower threshold settings, a maximum efficiency of about \SI{98}{\%} was measured at \SI{120}{V}.

\section{Conclusions}\label{sec:conclusions}
A readout system for the characterisation of the monolithic standalone matrices of the H35DEMO chip was fully developed at IFAE. The functionalities of the monolithic matrices of the chip were studied for the first time. The expected response to pulses and charged particles was obtained. The first beam test measurements of the monolithic CMOS matrix of a \SI{200}{\ohm cm} sample was carried out at CERN SPS showing a hit efficiency up to \SI{98}{\%} for a bias voltage of \SI{120}{V} obtained with a global tune of the chip. Further studies of the behaviour of chips with different resistivities before and after irradiation to the expected HL-LHC fluences are ongoing. 

\acknowledgments
This work was partially funded by: the Generalitat de Catalunya (AGAUR 2014 SGR 1177), the MINECO, Spanish Government, under grants FPA2015-69260-C3-2-R, FPA2015-69260-C3-3-R and SEV-2012-0234 (Severo Ochoa excellence programme); and the European Union's Horizon 2020 Research and Innovation programme under Grant Agreement no. 654168.
\FloatBarrier

\end{document}